\documentclass[%
reprint,
 amsmath,amssymb,
 aps,
prx,
]{revtex4-2}

\usepackage{color}
\usepackage{graphicx}% Include figure files
\usepackage{dcolumn}% Align table columns on decimal point
\usepackage{bm}% bold math
\usepackage{ulem}

\def\Vec#1{\boldsymbol #1}

\begin{document}

%\preprint{APS/123-QED}

\title{
Non-Abelian Quantum Turbulence in Spinor Bose-Einstein Condensates
}% Force line breaks with \\
%\thanks{Geometrical determination of internal degrees of freedom localized in vortex cores}%

\author{
    Michikazu Kobayashi
}
\email{kobayashi.michikazu@kochi-tech.ac.jp}
\affiliation{
    School of Engineering Science, Kochi University of Technology, Miyanoguchi 185, Tosayamada, Kami, Kochi 782-8502, Japan.
}
\author{
    Masahito Ueda
}
\affiliation{
    Department of Physics and Institute for Physics of Intelligence, The University of Tokyo, 7-3-1 Hongo, Bunkyo-ku, Tokyo 113-0033, Japan.
}
\affiliation{
    Fundamental Quantum Science Program (FQSP), TRIP Headquarters, RIKEN, Wako 351-0198, Japan.
}%

\date{\today}% It is always \today, today,

\begin{abstract}
We numerically investigate statistically steady quantum turbulence in the cyclic phase of a spin-2 spinor Bose-Einstein condensate (BEC) driven by large-scale energy injection via a divergence-free external velocity field.
In the full non-Abelian cyclic system, the mass-current spectrum exhibits an anomalous $k^{-7/3}$ scaling, whereas the spin-current spectrum approximately follows a $k^{-5/3}$ power law.
To isolate the role of non-Abelian vortex dynamics, we compare this system with an Abelian-restricted cyclic system sharing the same Hamiltonian parameters and driving conditions.
Strikingly, the Abelian restriction collapses both the mass- and spin-current spectra into a single $k^{-5/3}$ scaling, identical to that of a scalar BEC, despite the three distinct turbulent states maintaining nearly identical vortex-line densities.
A Helmholtz decomposition reveals that the mass- and spin-current spectra are predominantly incompressible and transverse, respectively, thereby confirming that the anomalous scaling cannot be attributed simply to compressible fluctuations or to the transverse nature of the energy injection.
Local spectral exponents further quantify these distinct scaling regimes.
The disappearance of the mass-spin spectral separation under the Abelian restriction, alongside the formation of a large-scale web of rung-connected non-Abelian vortices, provides compelling evidence that non-Abelian vortex dynamics fundamentally govern the organization of turbulent mass flow.
\end{abstract}

% \pacs{}

\maketitle

\section{Introduction}
Turbulence is a ubiquitous, strongly nonequilibrium phenomenon occurring in diverse macroscopic systems \cite{Frisch1995}.
While its instantaneous dynamics are chaotic and complex, turbulent flows frequently exhibit universal statistical properties.
A classic example is Kolmogorov's $-5/3$ law \cite{Kolmogorov1991a,Kolmogorov1991b,Batchelor1953}: within the inertial range, the kinetic-energy spectrum follows $E(k) \propto k^{-5/3}$ as energy cascades toward smaller length scales via a constant flux.

The statistical properties of turbulence are intimately linked to conserved quantities and their transfer across different length scales.
In two-dimensional incompressible turbulence, both kinetic energy and enstrophy - the spatial integral of the squared vorticity - are conserved in the inviscid limit.
Their simultaneous conservation gives rise to a dual-cascade phenomenology: energy cascades toward larger scales, whereas enstrophy transfers toward smaller scales, yielding the characteristic spectral scalings of $E(k) \propto k^{-5/3}$ and $E(k) \propto k^{-3}$, respectively \cite{Kraichnan1967}.
In three dimensions, helicity - defined as the spatial integral of the scalar product of fluid velocity and vorticity - constitutes yet another inviscid invariant alongside kinetic energy \cite{Thomson1868,Moffatt1969,Moffatt1992}.
In conventional three-dimensional turbulence, both energy and helicity generally cascade toward smaller scales \cite{Kraichnan1973,Borue1997,Chen2003a,Chen2003b,Gomez2004,Mininni2006}, with their joint transfer remaining compatible with the Kolmogorov $k^{-5/3}$ energy spectrum \cite{Brissaud1973,Krstulovic2009}.

Quantum turbulence is characterized by a spatially and dynamically complex network of quantized vortices \cite{Donnelly1991,TsubotaKasamatsuKobayashi2013,TsubotaKobayashiTakeuchi2013}.
In scalar superfluids, the incompressible kinetic-energy spectrum can exhibit the Kolmogorov $k^{-5/3}$ scaling at length scales larger than the characteristic intervortex spacing \cite{Kobayashi2005a,Kobayashi2005b,Navon2016,Muller2020}.
Spinor Bose-Einstein condensates (BECs) offer a qualitatively different platform, as their internal spin degrees of freedom accommodate both mass and spin dynamics.
Turbulent states in their systems have been shown to manifest unique spectral scalings associated with spin excitations and spin superflow \cite{Fujimoto2012,Fujimoto2014,Fujimoto2016}; accordingly, spin turbulence has been experimentally investigated in driven and quenched spinor condensates \cite{Kang2017,Hong2023,Jung2023,Kim2024}.
In particular, a $k^{-7/3}$ scaling has been reported for the spin-dependent interaction-energy spectrum in spin-1 turbulence \cite{Fujimoto2012,Fujimoto2014,Jung2023}, demonstrating that the additional internal degrees of freedom in a spinor superfluid can qualitatively alter its turbulent spectra.

An especially intriguing situation arises when the order-parameter manifold supports non-Abelian quantized vortices, whose topological charges are non-commutative.
Their non-Abelian nature has a direct consequence on vortex collisions: whereas Abelian vortices can reconnect and alter their geometric configuration \cite{Koplik1993,Koplik1996,Leadbeater2001}, collisions between non-Abelian vortices produce a rung vortex that bridges the colliding vortices \cite{Kobayashi2009,Mawson2015,Borgh2016}.
Spin-2 BECs host two distinct phases featuring such non-Abelian vortex topology: the biaxial-nematic phase and the cyclic phase \cite{Kawaguchi2012,Borgh2016,Xiao2022}.
At the mean-field level, however, the biaxial-nematic state belongs to an accidentally degenerate nematic manifold that is continuously connected to a uniaxial-nematic state.
The resulting quasi-Nambu-Goldstone modes enlarge the ground-state manifold, thereby destabilizing defects which would otherwise be topologically stable within the biaxial-nematic phase \cite{Uchino2010QN}.
While quantum fluctuations can lift this accidental degeneracy and stabilize either the uniaxial- or biaxial-nematic state \cite{Turner2007,Song2007,Uchino2010LHY}, incorporating such fluctuation-induced stabilization lies beyond the scope of the present mean-field framework.
We therefore focus on the cyclic phase, whose tetrahedral symmetry supports robust non-Abelian vortices already at the mean-field level.

Here, we numerically investigate statistically steady turbulence in the cyclic phase under large-scale energy injection.
To isolate the role of non-Abelian vortex dynamics, we compare the full cyclic system with both a scalar BEC and an Abelian-restricted cyclic system, the latter of which confines the dynamics to an Abelian invariant subspace within the same spin-2 model.
We find that the full non-Abelian cyclic system exhibits a clear separation between its mass- and spin-current spectra: the mass-current spectrum displays an anomalous $k^{-7/3}$ scaling, whereas the spin-current spectrum approximately follows a $k^{-5/3}$ power law.
In striking contrast, the Abelian-restricted system collapses both the mass- and spin-current spectra onto a single $k^{-5/3}$ scaling, mirroring the behavior of the scalar BEC.
A Helmholtz decomposition further reveals that these spectral laws persist in the predominantly transverse components of the corresponding currents.
These results demonstrate that the mass-spin spectral separation disappears under the Abelian restriction, providing compelling evidence that the anomalous $k^{-7/3}$ mass-current spectrum is intimately tied to the non-Abelian vortex dynamics of the full cyclic phase.

\section{Model}
We consider a spin-2 spinor BEC described by the Hamiltonian \cite{Kawaguchi2012}
\begin{align}
\begin{aligned}
&\mathcal{H} = \int d^3x \left\{ \frac{\hbar^2}{2 M} \sum_{m = -2}^2 |\nabla \psi_m|^2 \right. \\
&\phantom{\mathcal{H} = \int d^3x \quad} \left. \vphantom{\sum_{m = -2}^2} + \frac{g_0}{2} (\rho - \bar{\rho})^2 + \frac{g_1}{2} |\Vec{S}|^2 + \frac{g_2}{2} |A|^2 \right\},
\end{aligned}
\end{align}
where $\psi = (\psi_2, \psi_1, \psi_0, \psi_{-1}, \psi_{-2})^T$ is the spinor order parameter in the spin-2 Zeeman basis with $T$ being the transpose, $M$ is the atomic mass, and
\begin{align}
\begin{aligned}
& \rho = \sum_{m=-2}^2 |\psi_m|^2, \\
& \Vec{S} = \sum_{m,n=-2}^2 \psi_m^\ast [\hat{\Vec{S}}]_{m,n} \psi_n, \\
& A = \sum_{m=-2}^2 (-1)^m \psi_m \psi_{-m}.
\end{aligned}
\end{align}
Here $\bar{\rho}$ denotes the mean particle density, and $\hat{\Vec{S}}$ represents the vector of spin-2 matrices.
The conventional normalization factor $1 / \sqrt{5}$ in the spin-singlet pair amplitude is absorbed into the definition of $g_2$.
The density-interaction term is formulated relative to the uniform background density $\bar{\rho}$; for a fixed particle number, this differs from the conventional term proportional to $\rho^2$ only by a chemical-potential shift and an additive constant.

For $g_1>0$ and $g_2>0$, with $g_0>0$ ensuring stability against density fluctuations, the mean-field ground state at zero magnetic field belongs to the cyclic phase, characterized by $\Vec{S} = 0$ and $A = 0$ \cite{Makela2003,Makela2006,Zhou2006,Kawaguchi2012}.
A representative state of the cyclic phase is 
\begin{align}
\psi_{\rm cyclic} = \sqrt{\bar{\rho}} e^{i \phi} e^{- i \hat{\Vec{S}} \cdot \Vec{n} \theta} \left( \frac{i}{2}, 0, \frac{1}{\sqrt{2}}, 0, \frac{i}{2}\right)^T,
\end{align}
where $\phi\in[0, 2\pi/3)$ is the $U(1)$ phase, and $\Vec{n}$ and $\theta$ denote the unit vector along the rotation axis and the rotation angle about it, respectively.

In the full cyclic manifold, quantized vortices emerge as topological excitations that play a central role in quantum turbulence.
Due to the combined gauge-spin symmetry of the cyclic order parameter, the cyclic phase hosts vortices with fractional mass circulation.
In particular, the relevant vortex classes feature circulations $0$, $h/(3M)\equiv\kappa_{\rm cyclic}$, and $2\kappa_{\rm cyclic}$ \cite{Kobayashi2009,Kawaguchi2012}.
In the full non-Abelian cyclic turbulent state investigated below, vortices with zero and $\kappa_{\rm cyclic}$ circulations provide the dominant contributions, whereas those with $2\kappa_{\rm cyclic}$ are strongly suppressed due to their higher energetic cost.

To examine the role of non-Abelian vortex dynamics, we also investigate the restricted dynamics within a two-component subspace of the same spin-2 cyclic phase.
A representative state of this cyclic phase can be expressed as
\begin{align}
    \psi_{\rm cyclic}^{(2,-1)} = \sqrt{\frac{\bar{\rho}}{3}} \left(e^{i\phi_1},0,0,\sqrt{2} e^{i\phi_2},0\right)^T,
\end{align}
where $\phi_1$ and $\phi_2$ are independent phase angles representing combinations of the global gauge and spin rotations.
Motivated by this representation, we restrict the dynamics to
\begin{align}
    \psi_{\rm AR} = (\psi_2,0,0,\psi_{-1},0)^T. 
\end{align}
This subspace remains invariant under the Hamiltonian dynamics considered here: if the $m=1,0,-2$ components initially vanish, neither the interaction terms nor the spin-independent external energy injection can generate them.
Vortices in this restricted system are therefore characterized by the independent phase windings of $\psi_2$ and $\psi_{-1}$, whose topological charges commute.
We refer to this system as the Abelian-restricted (AR) cyclic system.
It serves as an ideal contrasting state, retaining the same spin-2 Hamiltonian and interaction parameters as the full cyclic system while excluding its non-Abelian vortex dynamics.

\section{Simulation of turbulence}
\begin{figure*}[tbh]
\includegraphics[width=0.99\linewidth]{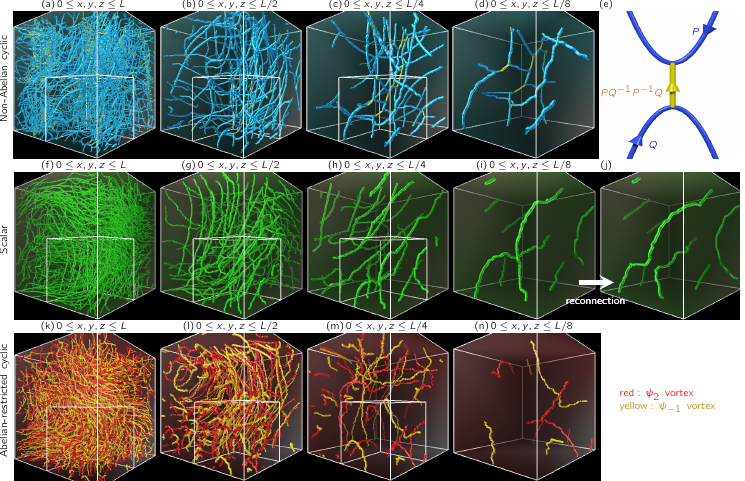}
\caption{
\label{fig:turbulence-image}
Snapshots of statistically steady quantum turbulence driven by large-scale energy injection via a divergence-free external velocity field for the non-Abelian cyclic spin-2 BEC (upper panels), the scalar BEC (middle panels), and the Abelian-restricted cyclic system (lower panels).
In the upper panels, blue and yellow curves represent vortices with circulations $\kappa_{\rm cyclic}=h/(3M)$ and $0$, respectively.
In the middle panels, green curves represent vortices of the scalar BEC with circulation $\kappa_{\rm scalar}=h/M$.
In the lower panels, red and yellow curves represent vortices in the $\psi_2$ and $\psi_{-1}$ components of the Abelian-restricted cyclic system, carrying circulations $\kappa_{\rm cyclic}$ and $2\kappa_{\rm cyclic}$, respectively.
Panels (a), (f), and (k) show the vortices in the entire simulation region, $0\leq x,y,z\leq L$.
The subsequent columns display progressively enlarged views: $0\leq x,y,z\leq L/2$ in (b), (g), and (l); $0\leq x,y,z\leq L/4$ in (c), (h), and (m); and $0\leq x,y,z\leq L/8$ in (d), (i), and (n).
In the non-Abelian cyclic system, vortices with circulation $\kappa_{\rm cyclic}$ are interconnected by zero-circulation rung vortices, forming a large-scale vortex web.
Rung formation is clearly visible in (d) and schematically illustrated in (e): when two non-Abelian vortices carrying non-commuting topological charges $P$ and $Q$ collide, they are linked by a rung vortex carrying the commutator charge $PQ^{-1}P^{-1}Q$.
In contrast, vortices in the scalar BEC undergo ordinary reconnection, as shown in (i) and schematically illustrated in (j).
The bottom panels show the turbulent vortex configurations in the Abelian-restricted cyclic system, where the dynamics are confined to the $(\psi_2,\psi_{-1})$ subspace; vortices in these two components carry commuting topological charges and are topologically allowed to pass through one another without forming rungs.
}
\end{figure*}
The dynamics of the system is governed by the coupled nonlinear Schr\"odinger equations;
\begin{align}
i\hbar\partial_t\psi_m
=
\frac{\delta\mathcal{H}}{\delta\psi_m^\ast}.
\end{align}
To achieve a statistically steady turbulent state, we introduce energy injection at large length scales, accompanied by dissipation at short length scales.
Dissipation is implemented by truncating the Fourier components of $\psi_m$ with wave numbers $k>k_{\rm c}$ during the time evolution.
Energy is injected by coupling the condensate to an externally imposed velocity field $\Vec{v}_{\rm ext}$ through the substitution
\begin{align}
\nabla
\rightarrow
\nabla-\frac{iM}{2\hbar}\Vec{v}_{\rm ext}.
\end{align}

We perform numerical simulations in a periodic cubic box discretized into $2048^3$ grid points with a spatial grid spacing $\Delta x=0.25\xi$.
Here,
\begin{align}
\xi=\frac{\hbar}{\sqrt{2Mg_0\bar{\rho}}}
\end{align}
is the healing length, which characterizes the typical size of a vortex core.
The system size is $L=512\xi$, and the cutoff wave number is set to $k_{\rm c}=2\pi/\xi$.
For the spin-2 systems, the interaction parameters are set to $g_1=g_2=0.5g_0$.

The external velocity field $\Vec{v}_{\rm ext}$ is constructed from a random long-wavelength vector field.
We first introduce an auxiliary field $\Vec{u}_{\rm ext}$ whose components are given by
\begin{align}
\begin{aligned}
\left(u_{\rm ext}\right)_i
=
\bar{v}
\sum_{0<|\Vec{n}|\leq n_{\rm c}}
\cos\left(
\frac{2\pi\Vec{n}\cdot\Vec{x}}{L}
+\theta_{\Vec{n},i}
\right),
\\
i=x,y,z,
\end{aligned}
\end{align}
where $\Vec{n}\in\mathbb{Z}^3$ and each phase $\theta_{\Vec{n},i}$ is independently and uniformly distributed over $[0,2\pi)$.

To obtain a divergence-free external velocity field, we perform a Helmholtz projection of $\Vec{u}_{\rm ext}$ in Fourier space.
For each nonzero wave vector $\Vec{k}=2\pi\Vec{n}/L$, the Fourier component is defined as
\begin{align}
\tilde{v}_{{\rm ext},i}(\Vec{k})
=
\sum_j
\left(
\delta_{ij}
-
\frac{k_i k_j}{|\Vec{k}|^2}
\right)
\tilde{u}_{{\rm ext},j}(\Vec{k}).
\end{align}
Consequently,
\begin{align}
\Vec{k}\cdot\tilde{\Vec{v}}_{\rm ext}(\Vec{k})=0,
\end{align}
or equivalently,
\begin{align}
\nabla\cdot\Vec{v}_{\rm ext}=0.
\end{align}
The real-space external velocity field $\Vec{v}_{\rm ext}$ is then obtained by the inverse Fourier transform.

For all simulations, we set $n_{\rm c}=2$, which corresponds to energy injection at large length scales.
In both the full non-Abelian and Abelian-restricted cyclic systems, we choose
$\bar{v}=0.4\sqrt{g_0\bar{\rho}/M}$.
Here, $\bar{v}$ specifies the amplitude of the auxiliary random field prior to the transverse projection.
Spatial derivatives are evaluated using a pseudo-spectral method, while time evolution is carried out via a fourth-order Runge-Kutta method with a time step of $\Delta t=0.001 \hbar / (g_0 \bar{\rho})$.
The random phases $\theta_{\Vec{n},i}$ are fixed throughout each realization, and the high-$k$ truncation is applied at every time step.

After a sufficiently long time evolution, each system reaches a statistically steady turbulent state characterized by a high vortex density.
Figures~\ref{fig:turbulence-image} (a)--(d) display the vortex configuration in the full non-Abelian cyclic system under the large-scale energy injection driven by the divergence-free external velocity field described above.
For comparison, we also simulate the Abelian-restricted cyclic system using the same spin-2 Hamiltonian parameters and the same value of $\bar v$ in the external-velocity-field construction.
The resulting vortex configuration is shown in Figs.~\ref{fig:turbulence-image} (k)--(n).

Furthermore, we perform simulations of a scalar BEC using the same energy injection mechanism and numerical setup.
For the scalar BEC, we use the corresponding single-component Gross-Pitaevskii model with the Hamiltonian
\begin{align*}
&\mathcal{H} = \int d^3x \left\{ \frac{\hbar^2}{2 M} |\nabla \psi|^2  + \frac{g_0}{2} (\rho - \bar{\rho})^2 \right\},
\end{align*}
and the same numerical procedure.
For the scalar BEC, we use $n_{\rm c}=2$ and $\bar{v}=0.8\sqrt{g_0\bar{\rho}/M}$, with $\bar v$ chosen so that the statistically steady state achieves approximately the same vortex-line density as in the two cyclic systems.
The resulting scalar turbulent state is shown in Figs.~\ref{fig:turbulence-image} (f)--(i).
Remarkably, the mean intervortex spacings of all three turbulent states agree within $2\%$, enabling a direct and reliable comparison of their spectral properties at comparable vortex densities.

The mass current of the spinor BEC is defined by
\begin{align}
\rho\Vec{v} = \bar{\kappa} \sum_{m=-2}^2 \operatorname{Im}\left(\psi_m^\ast \nabla \psi_m\right), \qquad
\bar{\kappa} \equiv \frac{\hbar}{M},
\end{align}
where $\Vec{v}$ is the superfluid velocity \cite{Kawaguchi2012}.
In terms of the density-weighted mass-current field
\begin{align}
\Vec{q}_{\rm mass} \equiv \sqrt{\rho} \Vec{v},
\end{align}
we define its spectrum $E_{\rm mass}(k)$ such that
\begin{align}
\int dk\: E_{\rm mass}(k)
= \frac{1}{2} \int d^3x\: |\Vec{q}_{\rm mass}|^2
\label{eq:kinetic-energy}.
\end{align}
The isotropic spectra are obtained by averaging the corresponding Fourier-space quadratic densities over spherical shells of constant $k=|\Vec{k}|$.
For a scalar BEC, the corresponding spectrum $E_{\rm scalar}(k)$ is given in Ref.~\cite{Nore1997}. % \cite{energy-divergence}
We further define the density-weighted spin-current fields as
\begin{align}
    \begin{aligned}
    \Vec{q}_{\rm spin}^{a}
    = \frac{\bar{\kappa}}{\sqrt{\rho}} \operatorname{Im}\left[\sum_{m,n=-2}^2\psi_m^\ast [\hat{S}_a]_{mn}\nabla\psi_n\right], \\
    a = x, y, z,
    \end{aligned}
\end{align}
where $\hat{S}_a$ denotes the spin-$2$ matrix along the $a$ direction.
The spin-current spectrum $E_{\rm spin}(k)$ is defined such that
\begin{align}
\int dk\: E_{\rm spin}(k)
= \frac{1}{2} \sum_{a = x, y, z} \int d^3x\: \left|\Vec{q}_{\rm spin}^a\right|^2.
\end{align}
The isotropic spectra are computed by averaging the corresponding Fourier-space quadratic densities over spherical shells of constant $k=|\Vec{k}|$.
In the following, the superscripts ${\rm NA}$ and ${\rm AR}$ distinguish spectra in the full non-Abelian cyclic and Abelian-restricted cyclic systems, respectively.

\begin{figure}[tbh]
    \centering
    (a) \\[-3pt]
    \includegraphics[width=0.95\linewidth]{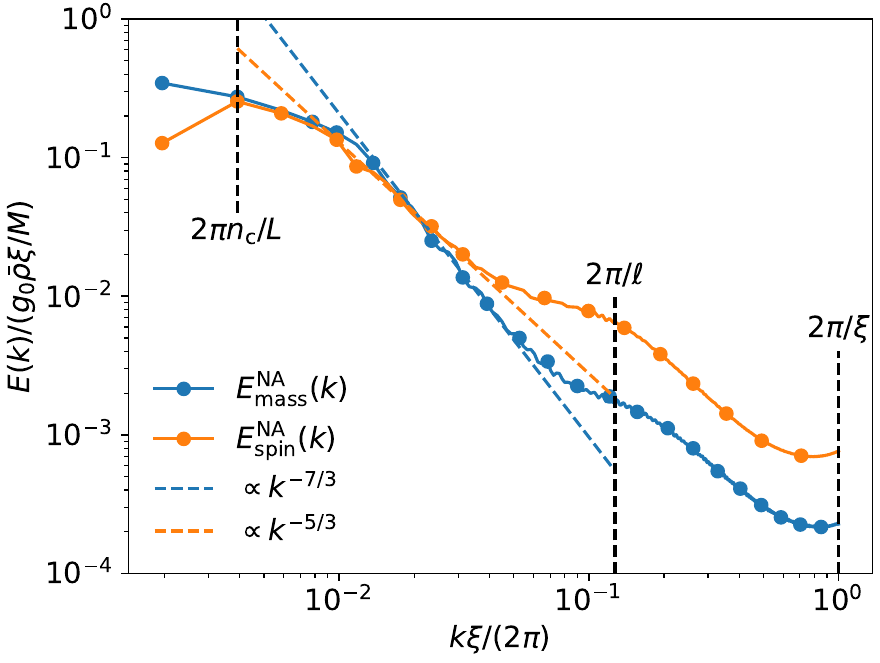} \\
    (b) \\[-3pt]
    \includegraphics[width=0.95\linewidth]{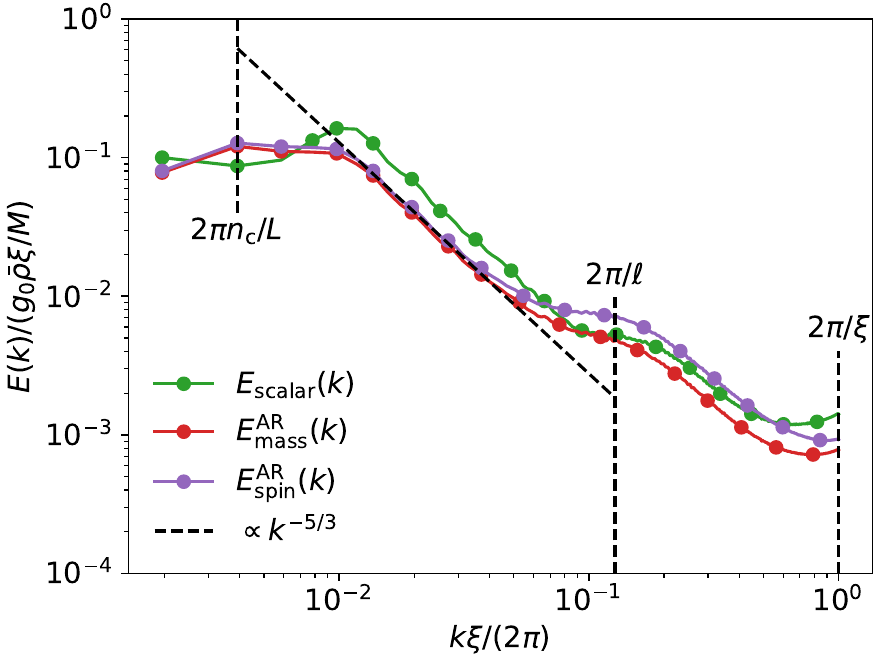}
\caption{
\label{fig:turbulence-spectrum}
Mass-current and spin-current spectra of (a) the full non-Abelian cyclic system,
$E_{\rm mass}^{\rm NA}(k)$ and $E_{\rm spin}^{\rm NA}(k)$,
and (b) the Abelian-restricted cyclic system,
$E_{\rm mass}^{\rm AR}(k)$ and $E_{\rm spin}^{\rm AR}(k)$,
plotted alongside the corresponding spectrum $E_{\rm scalar}(k)$ of the scalar BEC.
The systems are driven by large-scale energy injection via a divergence-free external velocity field.
Dashed lines indicate the $k^{-7/3}$ and $k^{-5/3}$ power laws.
Vertical lines mark key scales: the upper wave number of the energy-injection range $2\pi n_{\rm c}/L$, the characteristic intervortex wave number $2\pi/\ell$, and the healing-length wave number $2\pi/\xi$.
Here, $\ell=1/\sqrt{\rho_{\rm vortex}}$ represents the mean intervortex spacing, with $\rho_{\rm vortex}$ denoting the vortex-line density \cite{Kobayashi2016}.
For clarity, a single average value is used for $2\pi/\ell$ because the mean intervortex spacings of the three turbulent states differ by less than $2\%$.
Each spectrum is averaged over 200 statistically steady states generated from different realizations of the external random velocity field.
}
\end{figure}

As shown in Fig.~\ref{fig:turbulence-spectrum} (a), the mass-current spectrum of the full non-Abelian cyclic system exhibits a power-law scaling
\begin{align}
    E_{\rm mass}^{\rm NA}(k) \propto k^{-7/3}
\end{align}
spanning the range between the upper wave number of the energy-injection window and the characteristic intervortex wave number.
In contrast, the spin-current spectrum of the same turbulent state follows approximately
\begin{align}
    E_{\rm spin}^{\rm NA}(k) \propto k^{-5/3}.
\end{align}
A striking difference emerges when the dynamics are restricted to the Abelian subspace, as shown in Fig.~\ref{fig:turbulence-spectrum} (b).
In this case, both the mass- and spin-current spectra of the Abelian-restricted cyclic system approximately exhibit
\begin{align}
    E_{\rm mass}^{\rm AR}(k),\ 
    E_{\rm spin}^{\rm AR}(k)
    \propto k^{-5/3},
\end{align}
as does the corresponding spectrum of the scalar BEC,
\begin{align}
    E_{\rm scalar}(k) \propto k^{-5/3}.
\end{align}
Consequently, the separation between the mass- and spin-current spectral exponents is observed only in the full non-Abelian cyclic system.
In particular, the anomalous $k^{-7/3}$ mass-current scaling disappears when the dynamics are restricted to the Abelian subspace, even though both systems share the same interaction parameters and energy-injection conditions.
This distinction persists at nearly identical vortex-line densities.

To determine whether the distinct spectral scalings are associated with compressible or incompressible current fluctuations, we further analyze the mass and spin currents by decomposing them into their longitudinal and transverse components.
We perform a Helmholtz decomposition of the density-weighted current fields in Fourier space.
For the mass-current field, the longitudinal and transverse components are given by
\begin{align}
    \tilde{\Vec{q}}_{\rm mass}^c(\Vec{k})
    &= \frac{\Vec{k}\left[\Vec{k}\cdot \tilde{\Vec{q}}_{\rm mass}(\Vec{k})\right]}{|\Vec{k}|^2}, \\
    \tilde{\Vec{q}}_{\rm mass}^i(\Vec{k})
    &= \tilde{\Vec{q}}_{\rm mass}(\Vec{k}) - \tilde{\Vec{q}}_{\rm mass}^c(\Vec{k}),
\end{align}
where the superscripts $c$ and $i$ denote the compressible (longitudinal) and incompressible (transverse) components, respectively.
The mass currents of the full non-Abelian cyclic system, the Abelian-restricted cyclic system, and the scalar BEC are decomposed in a similar manner.

For the spin-current fields, we similarly define
\begin{align}
\tilde{\Vec{q}}_{\rm spin}^{a,L}(\Vec{k})
&= \frac{\Vec{k}\left[\Vec{k}\cdot\tilde{\Vec{q}}_{\rm spin}^a(\Vec{k})\right]}{|\Vec{k}|^2}, \\
\tilde{\Vec{q}}_{\rm spin}^{a,T}(\Vec{k})
&= \tilde{\Vec{q}}_{\rm spin}^a(\Vec{k}) - \tilde{\Vec{q}}_{\rm spin}^{a,L}(\Vec{k}),
\end{align}
where $L$ and $T$ denote the longitudinal and transverse components, respectively.
The corresponding spin-current spectra are obtained by summing over $a=x,y,z$.
For the Abelian-restricted system considered here, only the $a=z$ component is nonzero.

\begin{figure}[htb]
    \centering
    (a) \\
    \includegraphics[width = 0.95\linewidth]{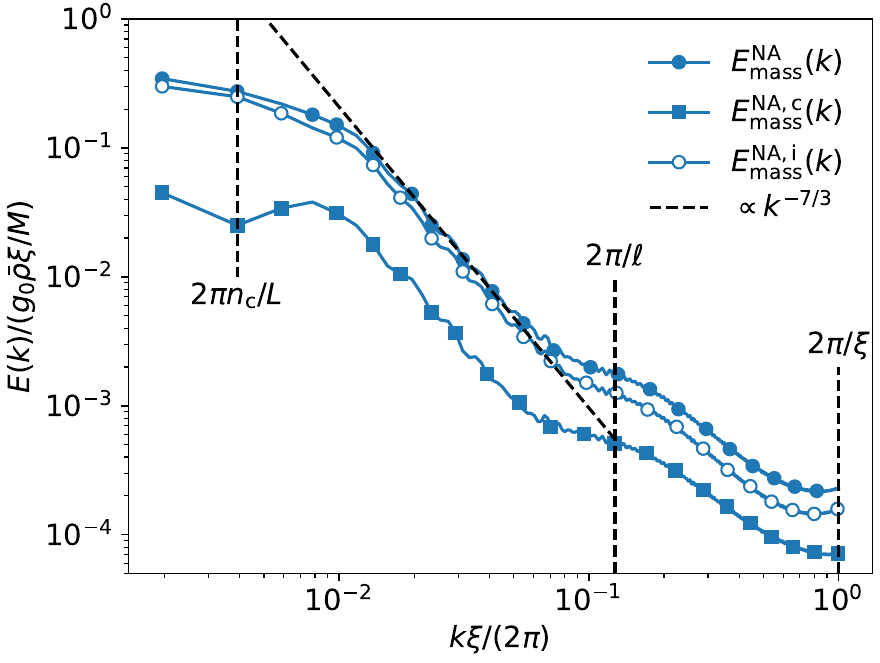} \\[5pt]
    (b) \\
    \includegraphics[width = 0.95\linewidth]{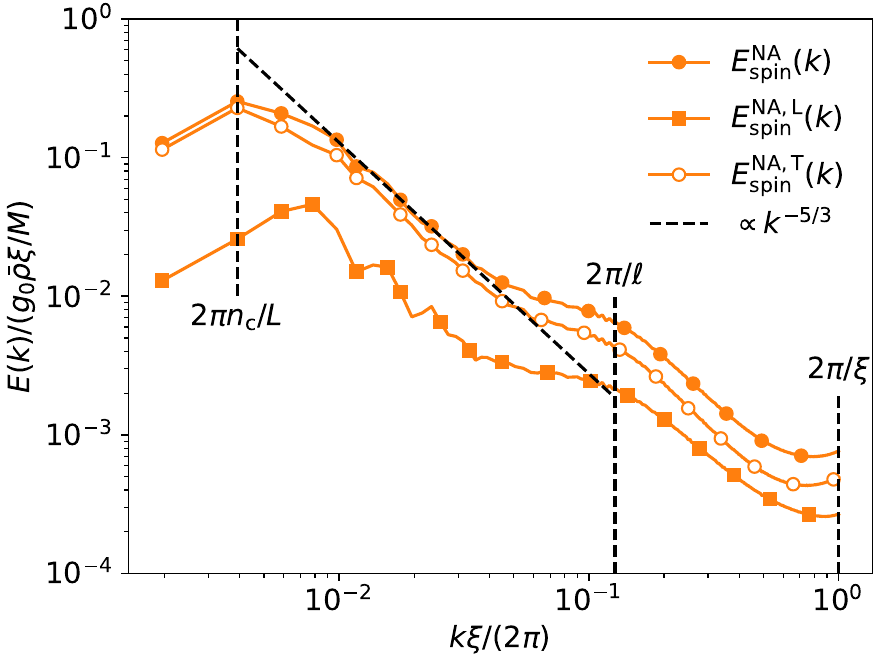}
    \caption{\label{fig:current-decomposition-1}
    Decomposition of current spectra in a full non-Abelian cyclic system driven by large-scale energy injection via a divergence-free external velocity field.
    (a) Total mass-current spectrum $E_{\rm mass}^{\rm NA}(k)$ and its compressible and incompressible components; $E_{\rm mass}^{{\rm NA},c}(k)$ and $E_{\rm mass}^{{\rm NA},i}(k)$.
    (b) Total spin-current spectrum $E_{\rm spin}^{\rm NA}(k)$ and its longitudinal and transverse components; $E_{\rm spin}^{{\rm NA},L}(k)$ and $E_{\rm spin}^{{\rm NA},T}(k)$.
    The vertical lines indicate the upper wave number of the energy-injection range $2\pi n_{\rm c}/L$, the characteristic intervortex wave number $2\pi/\ell$, and the healing-length wave number $2\pi/\xi$.
    }
\end{figure}

\begin{figure}[htb]
    \centering
    \includegraphics[width = 0.95\linewidth]{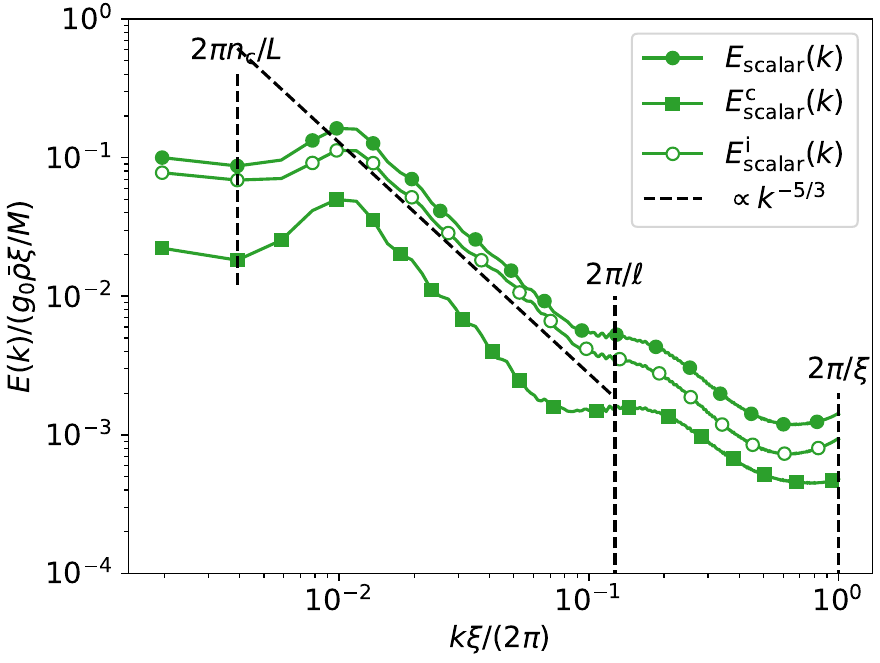}
    \caption{\label{fig:current-decomposition-2}
    Decomposition of the spectrum in a scalar BEC driven by large-scale energy injection via a divergence-free external velocity field.
    The total spectrum $E_{\rm scalar}(k)$ is decomposed into its compressible and incompressible components; $E_{\rm scalar}^{c}(k)$ and $E_{\rm scalar}^{i}(k)$.
    The vertical lines indicate the upper wave number of the energy-injection range $2\pi n_{\rm c}/L$, the characteristic intervortex wave number $2\pi/\ell$, and the healing-length wave number $2\pi/\xi$.
    }
\end{figure}

\begin{figure}[htb]
    \centering
    (a) \\
    \includegraphics[width = 0.95\linewidth]{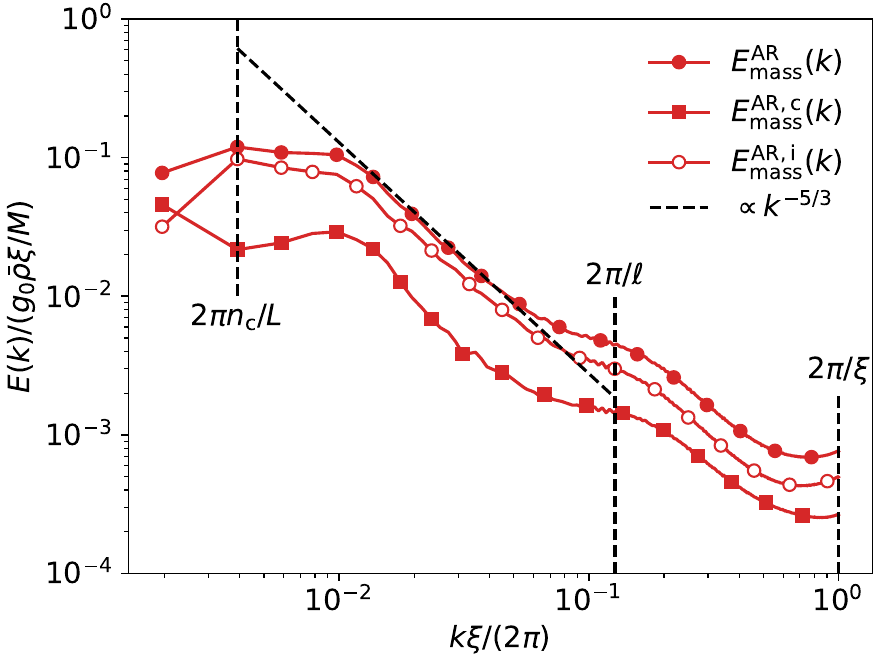} \\[5pt]
    (b) \\
    \includegraphics[width = 0.95\linewidth]{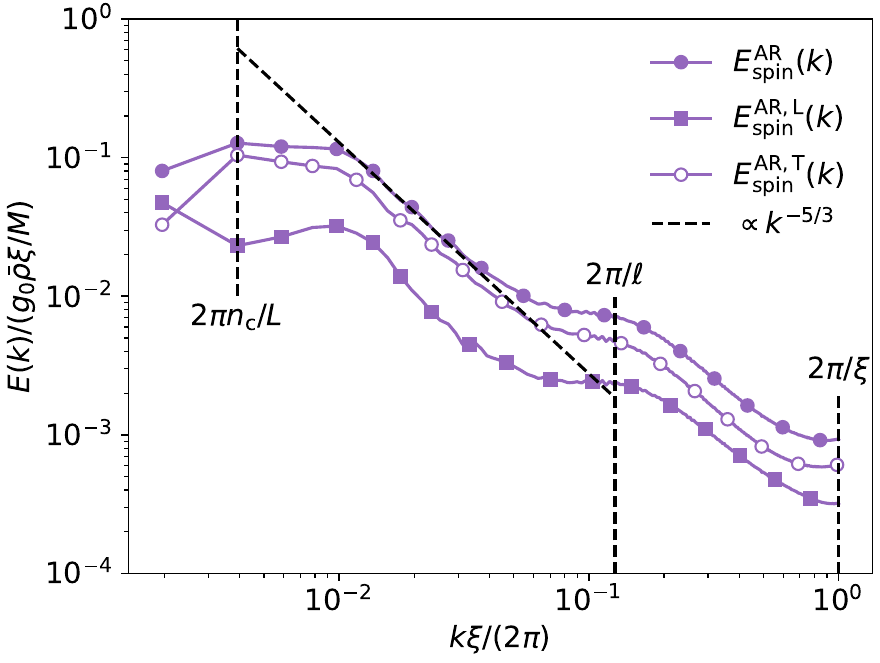}
    \caption{\label{fig:current-decomposition-3}
    Spectral decomposition of the current spectra in an Abelian-restricted cyclic system driven by large-scale energy injection via a divergence-free external velocity field.
    (a) Total mass-current spectrum $E_{\rm mass}^{\rm AR}(k)$ and its compressible and incompressible components; $E_{\rm mass}^{{\rm AR},c}(k)$ and $E_{\rm mass}^{{\rm AR},i}(k)$.
    (b) Total spin-current spectrum $E_{\rm spin}^{\rm AR}(k)$ and its longitudinal and transverse components; $E_{\rm spin}^{{\rm AR},L}(k)$ and $E_{\rm spin}^{{\rm AR},T}(k)$.
    The vertical lines indicate the upper wave number of the energy-injection range $2\pi n_{\rm c}/L$, the characteristic intervortex wave number $2\pi/\ell$, and the healing-length wave number $2\pi/\xi$.
    }
\end{figure}

Figures~\ref{fig:current-decomposition-1}--\ref{fig:current-decomposition-3}
show the Helmholtz decomposition of the current spectra.
For the full non-Abelian cyclic system, the incompressible mass-current spectrum
$E_{\rm mass}^{{\rm NA},i}(k)$ almost entirely overlaps with the total spectrum
$E_{\rm mass}^{\rm NA}(k)$ over the wave number range displaying a $k^{-7/3}$ scaling [Fig.~\ref{fig:current-decomposition-1} (a)].
Similarly, the spin current is dominated by its transverse component,
$E_{\rm spin}^{{\rm NA},T}(k)$, and both the transverse and total spin-current spectra approximately follow $k^{-5/3}$ scaling [Fig.~\ref{fig:current-decomposition-1} (b)] within the same non-Abelian turbulent state.

As shown in Fig.~\ref{fig:current-decomposition-2}, the scalar BEC is also dominated by its incompressible component.
Here, however, both the total and incompressible spectra approximately scale as $k^{-5/3}$.
The anomalous $k^{-7/3}$ behavior of the mass-current spectrum in the full cyclic system therefore cannot simply be attributed to compressible excitations.

A particularly insightful comparison is provided by the Abelian-restricted cyclic system shown in Fig.~\ref{fig:current-decomposition-3}.
As with the full non-Abelian system, its mass-current spectrum is dominated by the incompressible component, and its spin-current spectrum by the transverse component.
Nevertheless, both
$E_{\rm mass}^{{\rm AR},i}(k)$ and
$E_{\rm spin}^{{\rm AR},T}(k)$
approximately follow a $k^{-5/3}$ scaling, which is consistent with their corresponding total spectra.
Consequently, the $k^{-7/3}$ mass-current scaling and the associated separation between the mass- and spin-current exponents disappear when the cyclic dynamics is confined to the Abelian subspace.

The predominance of the incompressible mass-current components aligns with the energy injection via a divergence-free external velocity field.
In contrast, the spin current is not directly constrained by this transverse projection, even though its transverse component is also found to dominate.
Crucially, while predominantly transverse currents are observed in all three turbulent systems, the $k^{-7/3}$ scaling emerges only in the mass-current sector of the full non-Abelian cyclic system.
This anomalous scaling therefore can be attributed to neither compressible fluctuations nor the transverse nature of the energy injection.
Instead, its disappearance in the Abelian-restricted system provides compelling evidence that the mass-spin spectral separation is driven by the full non-Abelian cyclic dynamics.

To quantify the difference between the spectral scalings, we evaluate the local spectral exponent
\begin{align}
    \eta_\gamma^s(k)
    = - \frac{d\ln E_\gamma^s(k)}{d\ln k},
    \label{eq:def_eta_spinor}
\end{align}
where $\gamma = \{{\rm mass}, {\rm spin}\}$ and $s = \{{\rm NA}, {\rm AR}\}$.
For the scalar BEC, we similarly define $\eta_{\rm scalar}(k)$ by replacing $E_\gamma^s(k)$ in Eq. \eqref{eq:def_eta_spinor} with $E_{\rm scalar}(k)$.
Numerically, $\eta_\gamma^s(k)$ and $\eta_{\rm scalar}(k)$ are determined via local linear fitting of $\ln E_\gamma^s(k)$ and $\ln E_{\rm scalar}(k)$ against $\ln k$, respectively, over a seven-point window.

\begin{figure}[hbt]
    \includegraphics[width=0.95\linewidth]{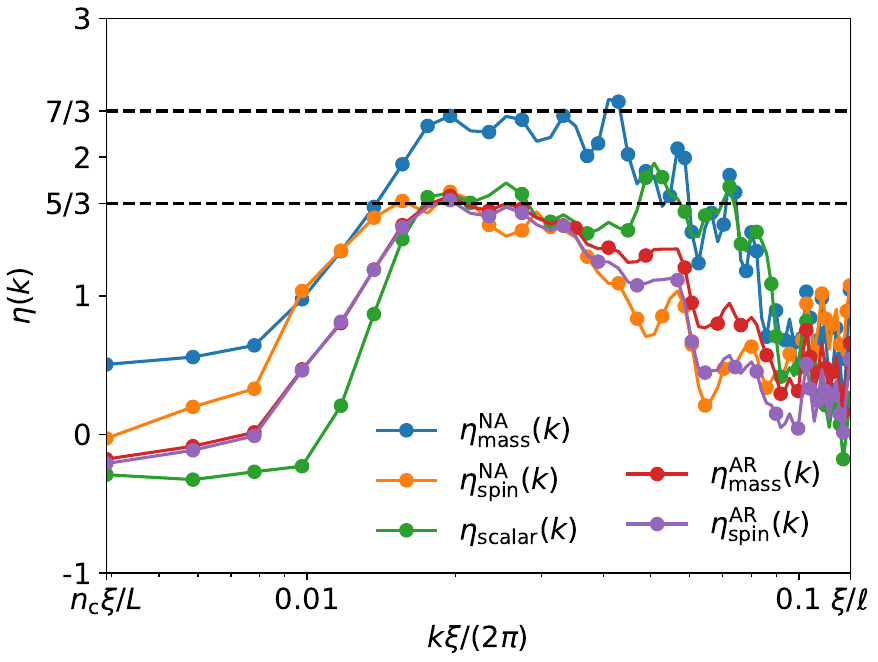}
    \caption{\label{fig:local-exponent}
    Local spectral exponents
    $\eta_{\rm mass}^{\rm NA}(k)$,
    $\eta_{\rm spin}^{\rm NA}(k)$,
    $\eta_{\rm mass}^{\rm AR}(k)$,
    $\eta_{\rm spin}^{\rm AR}(k)$,
    and $\eta_{\rm scalar}(k)$
    between the upper wave number of the energy-injection range $2\pi n_{\rm c}/L$
    and the characteristic intervortex wave number $2\pi/\ell$.
    The exponents are obtained from the mass- and spin-current spectra of the full non-Abelian cyclic system and the Abelian-restricted cyclic system, together with the corresponding spectrum of the scalar BEC.
    The horizontal dashed lines indicate $7/3$ and $5/3$.
    }
\end{figure}

As shown in Fig.~\ref{fig:local-exponent}, the five local spectral exponents display two distinct plateau-like regimes over a wide wave number range spanning from the energy-injection to the intervortex scales.
The mass-current exponent of the full non-Abelian cyclic system remains close to $\eta_{\rm mass}^{\rm NA}=7/3$.
In contrast, the remaining four exponents, $\eta_{\rm spin}^{\rm NA}$, $\eta_{\rm mass}^{\rm AR}$, $\eta_{\rm spin}^{\rm AR}$, and $\eta_{\rm scalar}$, stay close to $5/3$.
Deviations from these values become pronounced near both the energy-injection and intervortex scales, where the spectra depart from their approximate power-law behavior.

The local-exponent analysis therefore provides a complementary approach to quantitatively characterizing the spectral separation observed in the full non-Abelian cyclic system.
In particular, the anomalous mass-current exponent near $7/3$ changes to approximately $5/3$ when the dynamics are restricted to the Abelian subspace, whereas the spin-current exponent remains close to $5/3$ in both systems.
Combined with the Helmholtz decompositions shown in Figs.~\ref{fig:current-decomposition-1}--\ref{fig:current-decomposition-3}, these results demonstrate that the anomalous $k^{-7/3}$ scaling applies only to the predominantly incompressible mass-current sector of the full non-Abelian cyclic turbulent state.
This scaling is exhibited by neither the transverse spin-current spectra nor the incompressible mass-current spectrum of the Abelian-restricted cyclic system; instead, they remain close to $k^{-5/3}$, as does the scalar-BEC spectrum.

\section{Discussions and concluding remarks}
We have numerically investigated statistically steady quantum turbulence in the cyclic phase of a spin-2 spinor BEC under large-scale energy injection via a divergence-free external velocity field.
In the full non-Abelian cyclic system, the mass-current spectrum exhibits an anomalous $k^{-7/3}$ scaling, in marked contrast to an approximate $k^{-5/3}$ scaling in the spin-current spectrum.
However, when the dynamics are restricted to the Abelian subspace of the same spin-2 model under the same external-velocity-field conditions and with the same value of $\bar{v}$, both the mass- and spin-current spectra approximately scale as $k^{-5/3}$.
The scalar BEC exhibits a similar $k^{-5/3}$ spectrum.

The Helmholtz decomposition further reveals that the mass-current spectra are dominated by their incompressible components, whereas the spin-current spectra are predominantly transverse.
Although the predominance of the incompressible mass current is consistent with energy injection via a divergence-free external velocity field, the spin current is not directly constrained by this transverse projection.
Crucially, while predominantly transverse currents emerge in both the full non-Abelian and Abelian-restricted cyclic systems, the $k^{-7/3}$ scaling appears exclusively in the incompressible mass-current spectrum of the full non-Abelian system.
Analysis of the spectral exponents provides a complementary quantitative characterization of this distinction.
These findings demonstrate that the anomalous $k^{-7/3}$ scaling cannot be attributed simply to compressible fluctuations or to the transverse nature of the energy injection; furthermore, they reveal that the mass-spin spectral separation vanishes when the cyclic dynamics is restricted to the Abelian subspace.

The dynamical origin of the $k^{-7/3}$ mass-current spectrum remains an important open question.
Dimensional arguments suggest that such spectrum can arise when spectral transfer is controlled by a helicity flux \cite{Brissaud1973} -- a mechanism extensively investigated in quantum turbulence \cite{Leoni2016,Leoni2017,Villois2016,Scheeler2014}.
This observation makes helicity-related transfer a possible interpretation of the present spectrum.
However, the spectral scaling alone does not establish either the existence or the direction of a cascade.
In particular, the present analysis does not directly evaluate scale-by-scale energy or helicity fluxes.
Moreover, the quadratic measures defining $E_{\rm mass}$ and $E_{\rm spin}$ are not independently conserved quantities, so that a complete spectral budget must distinguish interscale transfer from conversion between different hydrodynamic sectors.
We therefore do not attribute the observed $k^{-7/3}$ spectrum to a pure helicity cascade at this stage.
Determining the detailed transfer mechanism underlying this scaling remains an open problem.

The present result should also be distinguished from the $k^{-7/3}$ scaling previously reported in spin turbulence of spin-1 BECs \cite{Fujimoto2012,Fujimoto2014,Jung2023}.
In these systems, the $k^{-7/3}$ law is associated with spin-related quantities, whereas in the full non-Abelian cyclic system studied here the spin-current spectrum approximately follows $k^{-5/3}$ scaling and the anomalous $k^{-7/3}$ scaling emerges in the mass-current sector.
A distinctive feature of the full cyclic dynamics is the non-Abelian collision of quantized vortices: vortices carrying non-commuting topological charges form rung vortices and do not undergo ordinary reconnection \cite{Kobayashi2009,Mawson2015,Borgh2016}.
As shown in Fig.~\ref{fig:turbulence-image}, repeated rung formation produces a large-scale connected web of non-Abelian vortices.
In contrast, in the Abelian-restricted cyclic system, the vortex charges commute, allowing vortices pass through each other without forming a rung.
Correspondingly, the anomalous $k^{-7/3}$ mass-current scaling and the mass-spin spectral separation observed in the full cyclic system both disappear under the Abelian restriction, with the mass- and spin-current spectra instead following approximately $k^{-5/3}$.
This concomitant shift in collision topology and spectral statistics strongly supports the link between non-Abelian vortex dynamics and large-scale mass flow beyond what a simple comparison with a scalar BEC alone.
Nevertheless, since the Abelian restriction also reduces the available dynamical degrees of freedom, it does not by itself establish a causal link between non-Abelian topology and the $k^{-7/3}$ scaling.

Our results thus reveal a characteristic separation between mass- and spin-current statistics in full non-Abelian cyclic turbulence -- a separation that disappears when the dynamics is restricted to an Abelian subspace of the same spin-2 model.
Whether the observed \(k^{-7/3}\) scaling represents a universal feature of non-Abelian quantum turbulence or is specific to the cyclic system remains an open question.
More broadly, the present results suggest that vortex topology may influence the organization of turbulent flow in multicomponent quantum fluids.

\begin{acknowledgments}
We would like to thank
M. E. Brachet, D. Proment, C. F. Barenghi, I. Danaila, and S.-i. Sasa for the helpful suggestions and comments.
The work of M. K. is supported by Grant-in-Aid for Scientific Research, Grants No. 23K22492, No. 24K00593, and No. 26K07020.
M. U. acknowledges the support by KAKENHI Grant No. JP22H01152 from the Japan Society for the Promotion of Science and the CREST program ``Quantum Frontiers" (Grant No. JPMJCR23I1) by the Japan Science and Technology Agency.
\end{acknowledgments}

%%%%%%%%%%%%%%%%%%%%%%%%%

%%%%%%%%%%%%%%%%%%%%%

\bibliography{non-Abelian-turbulence}% Produces the bibliography via BibTeX.

%%%%%%%%%%%%%%%%%%%%%%%%%%

\end{document}